 \documentclass[two column,sw]{agutex}
\usepackage{graphicx}
\usepackage{color,soul}
\usepackage[outdir=./]{epstopdf}
\usepackage{amssymb}
\usepackage{amsmath}

\authorrunninghead{SAVANI ET AL.}

\titlerunninghead{Bz prediction}

\begin{document}

%
%

\title{Predicting the magnetic vectors within coronal mass ejections arriving at Earth: 1. Initial Architecture}



%
%
%
\authors{N. P. Savani,\altaffilmark{1,2}  A. Vourlidas,\altaffilmark{1} A. Szabo,\altaffilmark{2} M. L. Mays, \altaffilmark{3,2} I. G. Richardson,\altaffilmark{4,2} B. J. Thompson,\altaffilmark{2} A. Pulkkinen,\altaffilmark{2} R. Evans, \altaffilmark{5} T. Nieves-Chinchilla,\altaffilmark{3,2}}

\altaffiltext{1}{Solar Section, Applied Physics Laboratory Johns Hopkins University, Laurel, MD USA}

\altaffiltext{2}{NASA, Goddard Space Flight Center, 8800 Greenbelt Rd, Greenbelt, MD 20771, USA}

\altaffiltext{3}{Catholic University of America, 620 Michigan Ave NE, Washington, DC 20064, USA.}

\altaffiltext{4}{University of Maryland, College Park, MD 20742, USA}

\altaffiltext{5}{College of Science, George Mason University, 4400 University Dr. Fairfax, VA 22030, USA }

%
%


\begin{abstract}
The process by which the Sun affects the terrestrial environment on short timescales is predominately driven by the amount of magnetic reconnection between the solar wind and Earth's magnetosphere. Reconnection occurs most efficiently when the solar wind magnetic field has a southward component. The most severe impacts are during the arrival of a coronal mass ejection (CME) when the magnetosphere is both compressed and magnetically connected to the heliospheric environment. Unfortunately, forecasting magnetic vectors within coronal mass ejections remains elusive. Here we report how, by combining a statistically robust helicity rule for a CME's solar origin with a simplified flux rope topology the magnetic vectors within the Earth-directed segment of a CME can be predicted. In order to test the validity of this proof-of-concept architecture for estimating the magnetic vectors within CMEs, a total of eight CME events (between 2010 and 2014) have been investigated. With a focus on the large false alarm of January 2014, this work highlights the importance of including the early evolutionary effects of a CME for forecasting purposes. The angular rotation in the predicted magnetic field closely follows the broad rotational structure seen within the in situ data. This time-varying field estimate is implemented into a process to quantitatively predict a time-varying Kp index that is described in detail in paper II. Future statistical work, quantifying the uncertainties in this process, may improve the more heuristic approach used by early forecasting systems.
\end{abstract}

%
%

%

\begin{article}

%
%

\section{Introduction}
CMEs are often observed to have twisted ``flux rope'' magnetic field structures \citep{liu2008,vourlidas2014}. If favorably oriented, these can lead to extended southward excursions of the interplanetary magnetic field (IMF) as the CME passes by, resulting in periods of enhanced reconnection on Earth's dayside and energy input into the magnetosphere. Draping of the IMF around the CME as it moves through the solar wind may also give rise to southward fields. In contrast, northward-directed fields inhibit reconnection, resulting in a weaker magnetospheric response \citep{dungey1961}. 
While prolonged southward fields are often observed without the presence of a clear structured transient, the additional plasma parameters often associated with CMEs usually make them the most geo-effective events \citep[e.g.,][]{tsurutani1997,zhang2014}.
Thus, inferring the direction of the flux rope fields inside a CME before it arrives at Earth would be a major advance in geomagnetic activity prediction.

In addition, the CME's initial configuration and its interaction with the inhomogeneous ambient solar wind can lead to deformations, rotations, and deflections of the magnetic field, which are difficult to quantify \citep[e.g.,][]{odstrcil1999,savani2010,nieves2012}. Distortions of CMEs have previously been observed by coronagraphs. However, their influence on the magnetic structure is difficult to estimate because the magnetically-dominated regions of CMEs appear as dark cavities within images, such as those seen by the STEREO spacecraft \citep{howardt2012b}. Therefore, a common approach to predicting magnetic vectors within a CME propagating towards Earth is to use solar observations as inputs into 3D computational simulations. Unfortunately, obtaining realistic magnetic field directions at Earth from such calculations is scientifically challenging and computationally intensive \citep{manchester2014}.

Thus, models used for routine CME forecasts by various space weather services do not include magnetic structures within the simulated CMEs \citep[e.g.,][]{Zheng2013, shiota2014}. For example, ENLIL models the propagation of CMEs from $\sim20$ solar radii (Rs) to beyond Earth at 215 Rs and includes the background solar wind magnetic field.  However, the CME is simplified to a high pressure plasma pulse with a size and propagation direction estimated from solar imagery \citep{Zheng2013}. CME arrival-time predictions from these models provide lead times of $\sim2-3$ days, and their accuracy has been well investigated \citep{taktakishvili2010, vrsnak2014, colaninno2013}. In contrast, the important magnetic vector information is only revealed when in situ measurements are made by spacecraft upstream of Earth at the first Lagrangian position (L1) $\sim1$ hour prior to the CME arriving at Earth, thereby severely limiting the lead time available for reliable, magnetic field-based, storm warnings.

Difficulties in observationally determining the magnetic profile of a CME arriving at Earth from only solar imagery predominately lie with several complex stages that change the initial solar configuration to the final topological structure at Earth. We suggest that for forecasting purposes, statistically significant predictions can be made by simplifying the complex behavior to a core set of parameters. In this paper, we highlight three key components of a proof of concept developed to improve the prediction of a storm's severity: 1. the use of the hemispheric helicity rule to provide a robust initial magnetic configuration at the Sun; 2. Define a `volume of influence' of the CME, within the heliosphere, for which the Earth's trajectory can be estimated; and 3. incorporating magnetic vectors from a simplified magnetic flux rope model to create a time-series upstream of Earth.

From the analysis of eight Earth-directed CMEs between 2010 and 2014, we conclude that the incorporation of magnetic field vectors in this way can significantly improve geomagnetic forecasts by providing a time-varying magnetic profile of the CME. The time varying magnetic profile is then incorporated with an experimental technique to create a time varying Kp index forecast that replicates the forecast deliverables by NOAA. Further details on the geomagnetic indices and their uncertainties that are borne from the vector estimates are described by Savani et al. (2015, herein referred to as Paper II).

\section{Event selection}
This article discusses the proof-of-concept architecture for estimating the magnetic vectors with the aid of a case study CME event that was released from the Sun on January 7th 2014 (see Figure \ref{sunA}). The recent release of this event allows for comparisons between the results described here and the current processes employed by real time space weather forecasters and their estimated geomagnetic indices (described further in Paper II).

A total of eight Earth-directed CME events between 2010 and 2014 were selected with three driving criteria: 1. Unambiguously define the solar source of the overlying field arcade from a single or double active region and possibly with an eruptive flare (see section \ref{SoIn} for more details); 2. A clear leading edge structure from multiple remote observations to unambiguously define the size and orientation of the CME morphology (see section \ref{Rem} for more details); and 3. a significant measurable effect by geomagnetic indices.

The eight events described in this paper were chosen from a CME list compiled by \cite{colaninno2013} and Patsourakos, S.(personal list), with details of each event displayed in Table 1. Further Earth-directed CME lists with more generic requirements have also been published \cite[e.g.,][]{richardson2010, mostl2014}.

The hemispheric solar source region of the CME is identified from solar observations. Figure \ref{sunA} displays a 171\AA{} image from the AIA instrument onboard the SDO spacecraft \citep{Lemen2012} taken at 20.14 UT on 7th January, 2014. This event has an inconclusive Earth-arrival time and in situ profile, and has been chosen to highlight the complexity in  forecasting processes. The uncertainty from this event stems from the predicted arrival time being approximately 24 hours earlier than when on-duty forecasters labeled the actual arrival from real-time L1 in situ data. Throughout this period, the solar wind plasma parameters displayed significantly lower velocities than were expected as well as missing a strong and distinctive magnetic field rotation of an obstacle.

\section{Solar Initiation} \label{SoIn}

The helicity and initial orientation of the magnetic flux rope structure within a CME are inferred from the ``Bothmer-Schwenn'' scheme. This relates the flux rope properties to sunspots, the solar cycle, and whether the CME originates on the northern or southern solar hemisphere \citep{bothmer1998}. The reliability of this solar hemispheric rule remains controversial. It was only in late-2013 when the probability of a CME's topology conforming to the hemispheric rule was re-confirmed to be $\geq$80$\%$ \citep{wang2013, hale1925}. Thus, the initial helicity and field structure of CMEs can be inferred from this scheme with a reliability that is likely to be $\sim$80$\%$.

Ordinarily, a CME is linked to a single active region where the standard Bothmer-Schwenn scheme should be applied. However in cases such as this January 2014 event, the magnetic loop structure before eruption has a leading negative polarity spanning over two active regions. Thus, a South-West-North, ``SWN'', flux rope field direction from southern hemisphere of solar cycle 23 under the Bothmer-Schwenn scheme is appropriate. This implies the CME has a right handed chirality. 

\cite{harra2007} highlighted the complexity of estimating the orientation of an interplanetary CME from simple solar observations. The work displayed that two CMEs released in November 2004 from a similar source location had drastically different final topologies. However, this can be reconciled with the Bothmer-Schwenn scheme if the different polarity of the active region's leading edge is taken into account.

In this article, we consider six simpler CMEs released from a single active region and examine whether it is possible to generate more reliable predictions of the field structure at 1 AU. We also investigate two more complicated cases where connected active regions are involved (September 27th, 2012 and January 7th 2014).

\section{Remotely Sensed Evolution} \label{Rem}

\begin{figure}
\noindent\includegraphics[width=44pc]{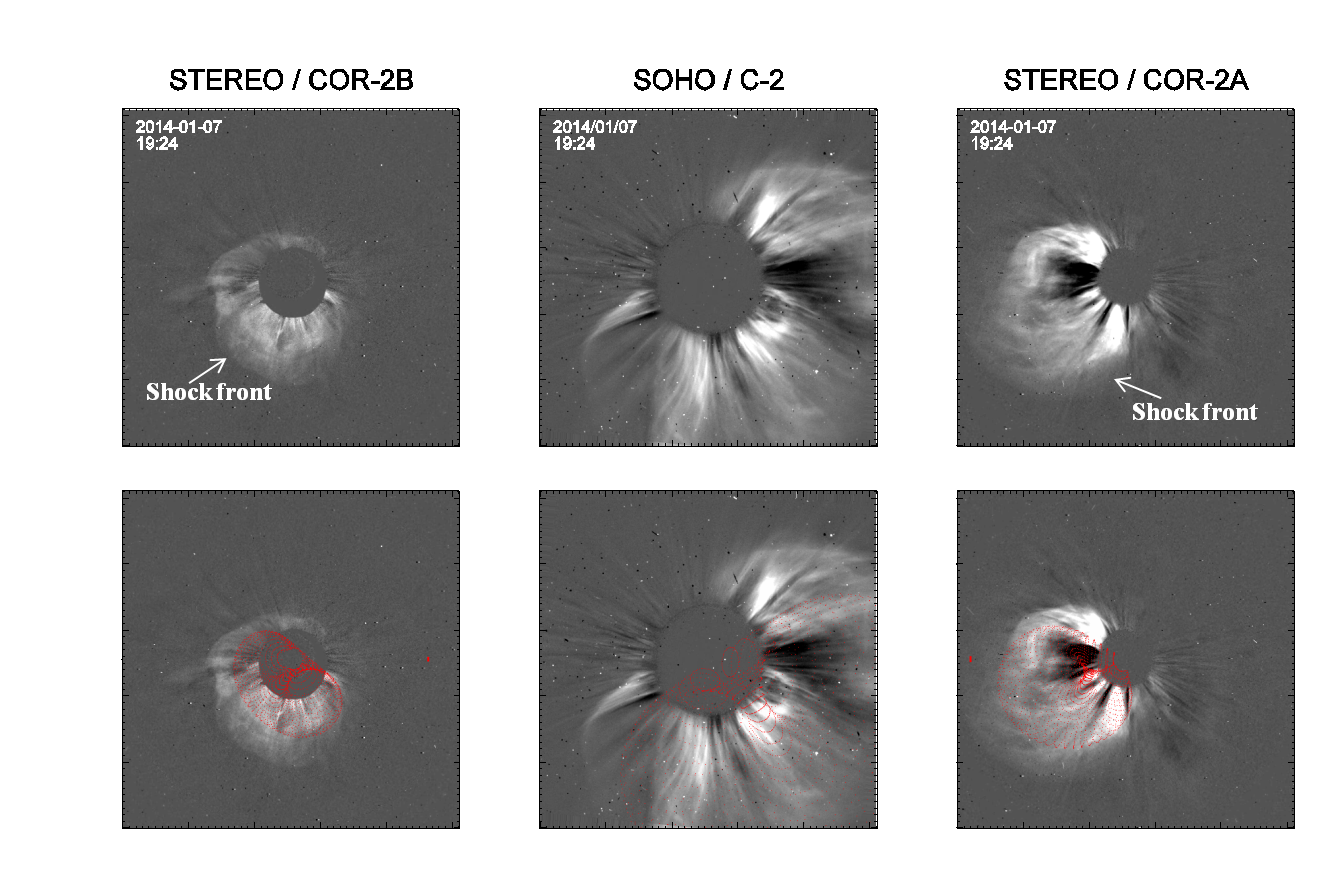}
	\caption{The evolution of the 07 January 2014 CME event from three vantage points. The graduated cylindrical shell model to estimate the topological structure of the event is shown in red. A transition layer ahead of the magnetic structure indicates the distance to the shock wave driven ahead.}
\label{gcsmodel}
\end{figure}

Since deflections, rotations and other interactions may occur during CME propagation to Earth, the initial Bothmer-Schwenn configuration is adjusted using coronagraphic data from the SOHO and STEREO missions \citep{brueckner1995,howard2008}. The final tilt and source region of the magnetic flux rope, after which radial propagation is assumed \citep{nieves2013}, is estimated using the graduated cylindrical shell (GCS) model \citep{thernisien2009}, when the CME reached $\sim 15$ Rs. Figure \ref{gcsmodel} displays images from the COR2 instruments onboard the two STEREO spacecraft (A and B) and the LASCO instrument onboard SOHO that are used to triangulate the CME structure. Where three well-separated observations exist, the GCS model provides relatively well-constrained estimates of the orientation and size of the CME without any ambiguity \citep{liu2010,rodriguez2011}. The GCS model may still be implemented without multi-point observations in the same way as various other cone-structure methods can be implemented. However in such cases, the possible degeneracy in the observational morphology limits all methods and thus highlights the difficulties in performing reliable forecasts.

\begin{figure}
\begin{center}
\includegraphics[width=44pc]{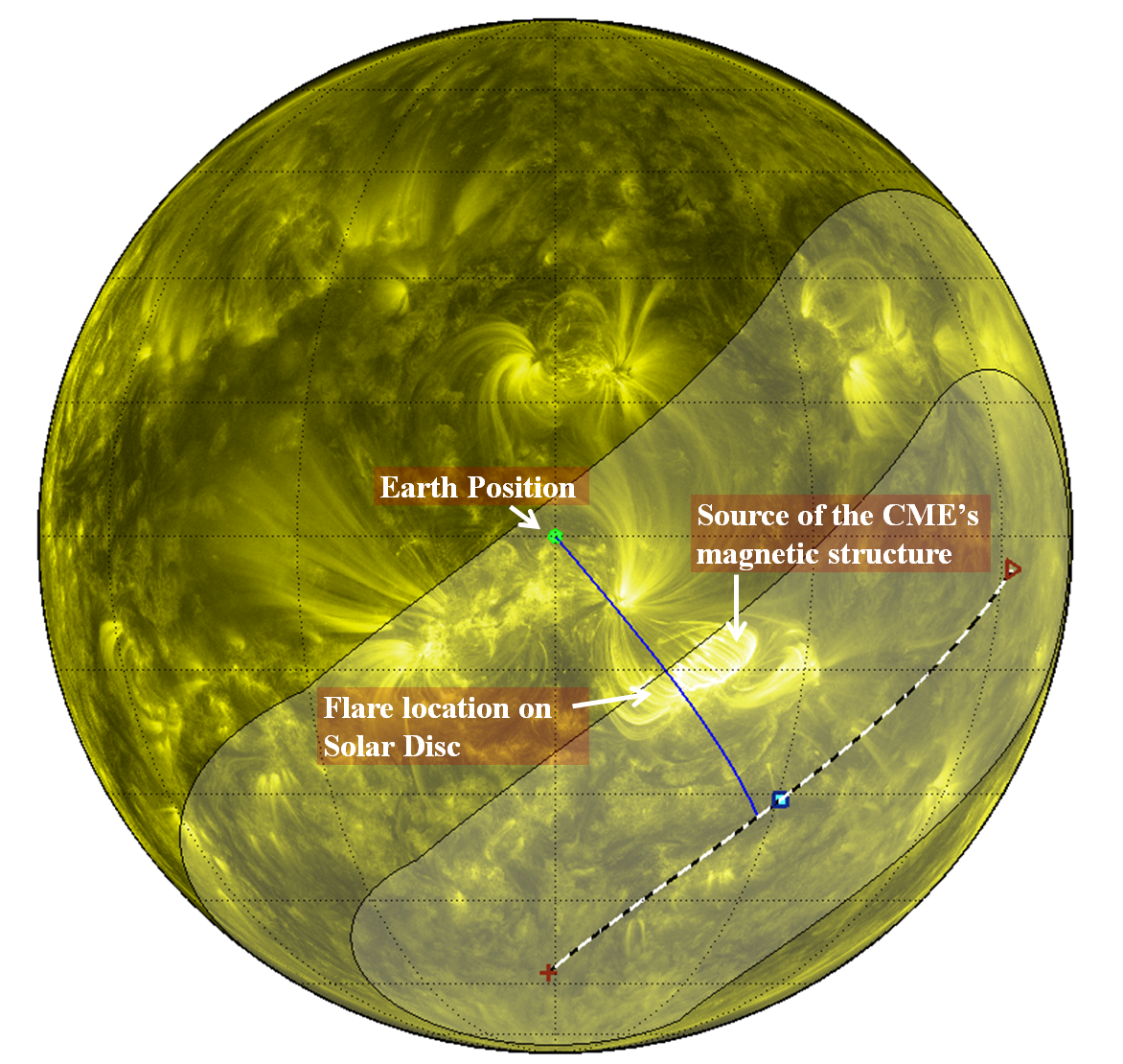}
	\caption{Solar source of the 7th January 2014 eruption. The Sun is shown as a 171Å (Fe IX/X) image from the AIA instrument onboard the SDO spacecraft taken at 20.14 UT before the propagation of the coronal mass ejection (CME). The extent of the axis of the CME magnetic structure is indicated by the dashed curve, displaying a southward deflection from the flare location. The center of the axis is shown with a blue square. The “volume of influence” onto the heliosphere from the CME is shaded, suggesting that the Earth only grazed the northern edge of the CME. The perpendicular distance of the Earth from the CME axis is shown in blue.}
\label{sunA}
\end{center}
\end{figure}

The outputs from the GCS model along with estimates of the average CME size \citep{yashiro2004} are used to create a ``volume of influence'' defined as the volume the CME is expected to traverse as it propagates through the heliosphere. The shaded region in Figure \ref{sunA} displays the projection of this ``volume of influence" onto the Sun, suggesting that the Earth grazed the northern edge of this case study event. The projected area is calculated from the `shadow' of the CME that is assumed to be cylindrical in shape with circular cross-section. Two parameters (flux rope axis length and flux rope width) are required to estimate the projected area. The axis length, shown as a dashed curve on Figure \ref{sunA}, is estimated from the half angle, $\alpha_{haw}$, given by half the angular width of the CME in a direction parallel to the GCS model axis. The projected width of the CME transverse to this axis is assumed equal to the average CME width, as found from statistical studies \citep{nieves2013}. 

Any uncertainty in the inferred CME orientation is likely to have only a minimal effect on the predicted magnetic field vectors since it is likely to be eclipsed by the larger uncertainties arising from estimating the magnetic field strength and the assumption of a symmetric cylindrical flux rope, as explained below. Further testing of these assumptions are addressed in paper II and are tested relative to predicted estimates of Kp. The coronagraphic images show how the coronal magnetic loops seen in SDO have been deflected to the south west (Figure \ref{sunA}). Here, we use coronagraphic imagery to estimate the final CME radial trajectory but future work could attempt to increase prediction lead times by, for example, incorporating CME deflections by coronal holes \citep{cremades2004, makela2013}. 

\begin{figure}
\begin{center}
\includegraphics[width=25pc]{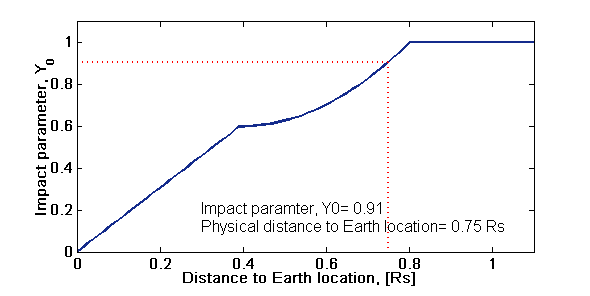}
	\caption{An empirical model comparing the CME axis to Earth distance with a theoretical magnetic structure length. The January 7, 2014 event is shown to graze the outer edge of the magnetic structure, with a normalised impact parameter of Y0=0.91.}
	\label{ImpParam}
\end{center}
\end{figure}

The shortest (perpendicular) distance between the Earth's projected location and the flux rope axis is indicated on Figure \ref{sunA} with a blue curve. Normalizing this to the total perpendicular distance to the flux rope outer edge (flux rope radius) gives a quantity that is correlated with the impact parameter (Y0) which is a key parameter for in situ flux rope modelers. The theoretical model of the impact parameter used in this study is displayed in Figure \ref{ImpParam}; the Earth's projected arc distance is displayed in solar radii. This theoretical function is justified by using a simple linear correlation between the Earth's distance and Y0 for the inner core region surrounding the flux rope axis (inner highlighted area in Figure \ref{sunA}). The outer area is correlated with a trigonometric sine function and is designed to physically represent the distortions to the idealized flux rope that occur during propagation as well as possible draping of the surrounding solar wind magnetic field outside the actual flux rope structure. These distortions to the flux rope are sometimes termed `pancaking' \citep{riley2004a,savani2011a} with recent studies suggesting the inner core of a CME is likely to maintain a quasi-cylindrical structure while the outer structure may become severely deformed by the ambient medium \citep{demoulin2009, savani2013b}. 

\section{In Situ Flux Rope}

To generate an estimate time-series of the magnetic vector direction passing over a fixed point such as L1, we must employ a methodology to create a 1-D (spacecraft) trajectory through a theoretical structure, and to define the start time of the object at this fixed point.  

\subsection{Time of Arrival}
To improve the time of arrival prediction of a CME is beyond the scope of this work, and several advances on this topic have been performed. Currently there are several procedures to calculate the speeds of remotely-observed CMEs, quantify their deceleration, and forecast their speeds upstream of Earth at L1 \citep[see further literature within, e.g.,][]{owens2004,colaninno2013,tucker2015}. We choose to assume a simple average of the measured CME speed close to the Sun as determined by the NOAA Space Weather Prediction Center (SWPC) and the predicted speed at Earth.  In the case of the 7th January, 2014 event, this gives a CME speed of 1300 km/s and a predicted arrival time of 8th January, 21.45 UT. By combining this information with the flux rope model described below, a time-series of magnetic vectors is created. 

In order to compare the accuracy of the modeled magnetic vector time-series with data and test the technique in the `research domain', we manually adjust the arrival time of the model fit to the best guess estimate within the L1 data. The field rotations between the model estimate and data were then manually inspected. 

However, for a readily implementable process for estimating the magnetic vectors in advance, different forecasting systems can simply employ their best estimate of the arrival time.

\subsection{Flux Rope Model}
The configuration of the magnetic flux rope is calculated by assuming a constant-alpha force-free (CAFF) flux rope \citep[and references therein]{burlaga1988,zurbuchen2006} and a cylindrical geometry locally around the Earth's predicted trajectory through the CME. Previously, triangulation of the CME direction from remote sensing have provided adequate information as to the expected structure arriving at L1 \citep{liu2010b}. However a Grad-Shafranov reconstruction technique used by \cite{liu2010b} would not be appropriate in creating a model to estimate the structure. Future work may consider implementing a more complex model that better represents the distortions occurring to a CME at L1 \citep[e.g.,][]{marubashi1997,hidalgo2002,owens2006}

The magnetic vectors generated along the Earth's trajectory from a CAFF flux rope model is created from the MHD momentum equation under magnetostatic equilibrium; which can be reduced to $\textbf{j}=\alpha\textbf{B}$. A solution to this equation can be used to generate a cylindrical magnetic flux rope with circular cross section, with the components of the magnetic field vector expressed by Bessel functions, and $\alpha$ commonly set to 2.41 \citep{savani2013a}. Future work can consider reducing $\alpha$ as a simple solution to potential flux erosion occurring to the CME during propagation \citep{ruffenach2012}.

\begin{figure}
\includegraphics[width=25pc]{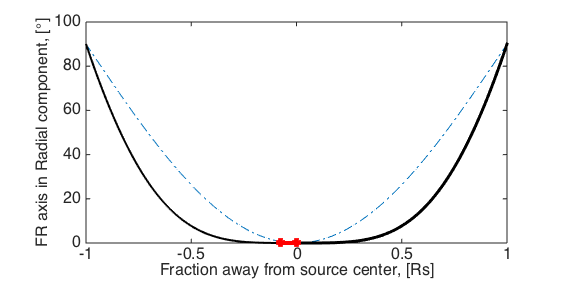}
	\caption{An empirical model to estimate the radial component of the flux rope axis direction ($\lambda$). The relative distance of Earth impact and the flux rope nose is shown as the red line. The flux rope axis direction is assumed to be perpedicular to the radial at the nose and parallel to the radial ($90^\circ$) at both footpoints.}
	\label{RadComp}
\end{figure}

The projected axis onto a 2D plane of the CME is provided by a single angle orientation ($\phi$) estimated from GCS model. However, the component of the flux rope axis parallel to the radial direction is estimated theoretically, by measuring the shortest distance between the Earth trajectory through the CME away from the CME nose, Ln. In practice, this was performed by measuring half the length of the flux rope axis ($R_{ax}$) and the length between the flux rope axis center and the position where the Earth perpendicular position (Figure \ref{sunA}, blue curve) meets the flux rope axis ($D_E$); thereby defining, $L_n \equiv D_E/R_{ax}$. $L_n = 0$ represents the case where the CME nose is propagating directly towards Earth, and there is no radial contribution to the flux rope axis. However when $L_n =1$, the flux rope axis is entirely radial in direction, as might be the case when the Earth's trajectory is along a CME leg. Figure \ref{RadComp} displays how the radial contribution to the axis vector is estimated from an angular value ($\lambda$) that varies between $0^{\circ}$ (CME nose) and $90^{\circ}$ (CME leg) in a scheme similar to that expressed by \cite{janvier2013}. Both $\phi$ and $\lambda$ are used to create a 3D flux rope axis direction. 

The magnitude of the magnetic field along the central flux rope axis is assumed in this case study to be 18.0 nT. This is calculated by assuming the maximum estimated magnetic field strength within the plasma pile-up region simulated by the WSA-ENLIL+Cone model (10.3 nT) corresponds to the magnetic field strength at closest approach within the flux rope structure. The impact parameter obtained using the `volume of influence' (set at 0.91 for the January 2014 event) is then used to estimate the maximum field strength along the central flux rope axis. In effect, this technique estimates the $\mid\textbf{B}\mid$ of a CME from a correlation of the inner heliospheric CME velocity and a simulated background solar wind field strength driven by magnetograms.

\begin{figure}
\includegraphics[width=25pc]{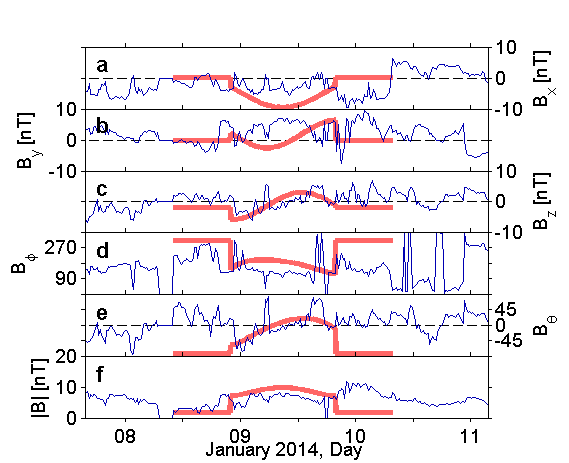}
	\caption{Magnetic vectors from the L1 vantage point upstream of Earth for the arrival of the CME. The magnetic field from the OMNIWeb dataset is shown in GSE components (a-c), and spherical coordinates (d-f). The red curves overlaid represent the forecasted magnetic vectors at Earth.}
\label{MagVec}
\end{figure}

The flux rope axis direction, chirality, magnetic field magnitude and impact parameter provide a complete set of parameters to generate a time series of magnetic vectors along a theoretical Earth trajectory (Figure \ref{MagVec} red curves).

\subsection{Magnetic Field Strength} 
The field strength is inferred from a model currently used for forecasting by CCMC, so this method could be implemented using existing forecasting capabilities. In the future, other methods whose uncertainties have not yet been statistically quantified might be used, for example estimating the poloidal and total flux content of a CME from flare ribbon brightening \citep{longcope2007}; flux rope speed and poloidal flux injection to estimate field strength \citep{kunkel2010}; using radio emissions from the CME core \citep{tun2013}; and using the shock stand-off distance from remote observations which has recently shown the possibility of estimating the field strength upstream of a CME \citep{savani2011b, poomvises2012, savani2012b}.

Considering the final focus of estimating the magnetic vectors is to predict the terrestrial effects with quantifiable uncertainty, the uncertainty in the predicted Kp index was estimated by varying the field strength over the range $\mid\textbf{B}\mid=18.0_{-1\sigma}^{+2\sigma}$, where $\sigma$ = 6.9nT \citep{lepping2006}. The uncertainty in field strength represents the statistical average from 82 flux rope fittings estimated between 1995-2003. The magnetic vectors were recalculated for each field strength and used to drive estimates of Kp which are described in more detail in Paper II.

\section{Results} \label{res}
\begin{figure}
	\includegraphics[width=30pc]{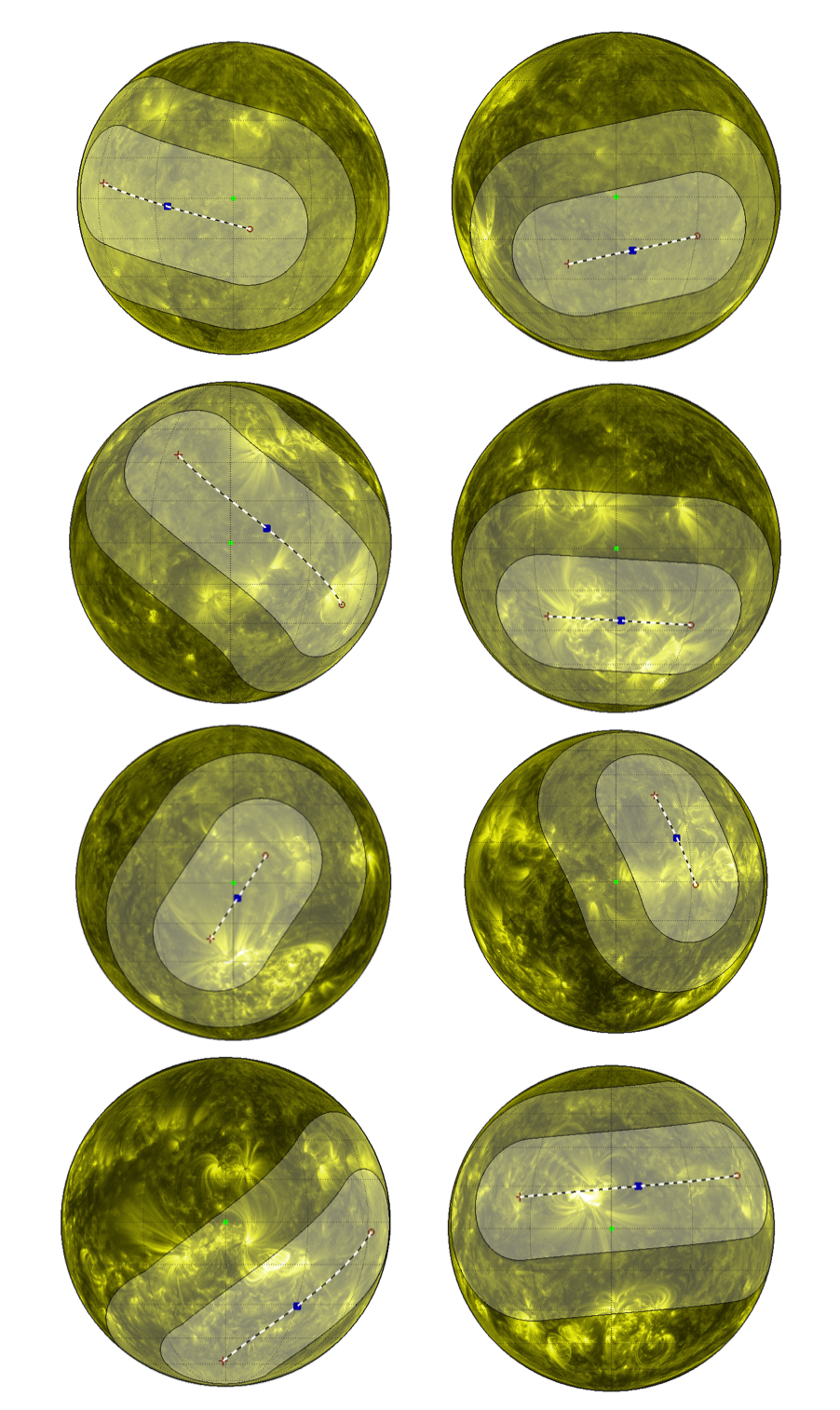} 
	\caption{Solar source of 8 CME eruptions between 2010 and 2014. The Sun is shown as a 171Å (Fe IX/X) image from the AIA instrument onboard the SDO spacecraft and in the same format as Figure \ref{sunA}.}
	\label{solarsurvey}
\end{figure}

In order to test the validity of this proof-of-concept architecture for estimating the magnetic vectors within CMEs, a total of eight CME events have been investigated. By using the same technique as Figure \ref{sunA}, the solar disc in 171\AA $\,$ AIA and projected CME `volume of influence' for these events between 2010 and 2014 are displayed Figure \ref{solarsurvey}. Their predicted magnetic vectors are displayed in Figure \ref{Bsurvey}, along with the measured L1 in situ data. Spherical coordinates are used to display the magnetic field rotation over the Cartesian system as the orientation components remain independent of the magnetic field strength component. The angular rotation ($B_\phi$ and $B_\theta$) in the predicted magnetic field closely follows the broad rotational structure seen within the in situ data, with a negative $B_\theta$ indicating a southward magnetic field excursion. The deviations in results between the estimated and measured values are discussed below.

For the events investigated, it has been noticed that if the overlying magnetic field arcade displayed in solar imagery (e.g., within 171\AA $\,$ AIA) traverses two active regions in close proximity, an adjustment to the standard scheme is required. In particular, if the solar arcade is between two active regions, the leading polarity is reversed and the initial magnetic structure is defined by the Bothmer-Schwenn scheme from the previous solar cycle
. The scenario of this more complex behavior is shown in the case study event of this article (January 2014, panel G in Figures \ref{solarsurvey} and \ref{Bsurvey}), as well as in an event on September 2012 (Figure \ref{solarsurvey} and \ref{Bsurvey}, panel F). Therefore, we suggest that the ubiquitous use of the Bothmer-Schwenn scheme with a simplistic flux rope model is capable of generating a zeroth-order characterization of the rotating magnetic field topology with a flux rope CME. 

Figure \ref{Bsurvey} also illustrates the limitations of a symmetrical flux rope model, which is frequently highlighted by a variety of in situ models, in that the model field strength are by definition stronger near the center of the flux rope whereas the observed fields occasionally deviate from this pattern. As an example, panels (a), (e) and (f) display the strongest field near the CME leading edge or sheath, which sometimes occurs when a fast CME compresses against the solar wind ahead.

The rotating nature of the magnetic field's southward excursion has important consequences for improving start time predictions of significant values in Kp index or aid strength estimates of the Dst storm onset. Panels (a), (b) of Figure \ref{Bsurvey} show examples of an initial prolonged northward magnetic field component and thereby would predict a delayed start of large Kp values (see Paper II for more details).

There are also processes that can influence the accuracy of the predicted fields, in particular the interaction of CMEs during passage from the Sun to the Earth \citep{shen2012}. As an example, two CME events launched in quick succession were detected as a single strong event within in situ data and displayed in panel C of Figure \ref{Bsurvey}. In the interim, an experienced observer may be able to manually adjust the computational models in response to such unusual situations in a heuristic manner used by forecasters.

The estimation of the impact parameter (i.e., perpendicular to the flux rope axis) is an important variable in influencing the predicted magnetic vector. This parameter affects the estimated peak magnetic field strength as well the expected angular change in the field rotation. The influence on total field rotation goes from observing a maximum $180^\circ$ rotation to a minimum of $0^\circ$ between a trajectory through the core and edge, respectively. As an example, the predicted vectors for panel C and H in Figure \ref{Bsurvey} display a significantly larger variation in field rotation than was observed.

For the case of the January 2014 event, draping of the surrounding solar wind magnetic field is likely to account for significant portion of the measured terrestrial disturbance due to Earth's trajectory through the outer northern edge. As a first principle, the field rotation from a draped magnetic field as measured from a 1-D spacecraft trajectory can be modeled with the minimal rotations created from a large impact parameter modeled flux rope described below. For cases as extreme as this, a forecast system that generates a subtle field rotation may be considered more appropriate than generating a `missing-Earth' scenario, but extensive statistical analysis will be required to minimize uncertainty for such cases.

Statistically, a spacecraft should have no relationship with the CME trajectory, and the frequency distribution of CME versus the spacecraft distance is expected to be approximately uniform. This has not always been observed with in situ detectors \citep{lepping2010}, but this is likely due to the trajectory being outside their core flux rope behaviour \citep{demoulin2013}. Therefore a split behaviour of the impact parameter, used in this work, between the central core and the outer regions is an appropriate choice. A common uncertainty in impact parameter from various models is considered as approximately $\pm10\%$ \citep{alhaddad2012}. Therefore in Paper II, changes to the impact parameter over the uncertainty range are used to create an ensemble of predicted vectors in order to investigate their consequences on the Kp index.

The estimated magnetic vectors from the CME is quasi-invariant to any trajectory variations that are parallel to the flux rope axis. This is because the simplistic model is an axis-symmetric cylinder. The small changes to the estimated vectors that does occur is a result of small adjustments to the radial component of the CME axis direction. The estimated vectors change rapidly once the predicted trajectory approaches the legs of the flux rope axis as the influence of the radial component is highly non-linear. Under such situations, detection of a CME with in situ data usually becomes inconclusive \citep{owens2012} and therefore unlikely to have a major impact for the purposes of predicting large Kp values at Earth.


\section{Discussion, Conclusions and Future Work}

This article displays a reliable mechanism by which magnetic vectors can be forecasted. The current process lays the organizational structure that is based on remote sensing and empirical relationships. The example January 2014 event is severely deflected away from the Sun-Earth line and thus highlights the importance of including evolutionary estimates of CMEs from remote sensing when attempting to provide reliable forecasts (as previously suggested by \cite{mcallister2001}). Also, to improve the reliability of the magnetic vector forecast, the initial topological structure determined by the Bothmer-Schwenn scheme must be adjusted for cases where the overlying field arcade clearly traverses two active regions.

While the current process lays the organizational structure, in its current format, the concept has not yet been statistically proven to be more beneficial at helping estimate the geo-effectiveness of an Earth-directed CME. For this, Paper II describes a first approach to create an ensemble of magnetic vector predictions that are used to create predicted geomagnetic indices, (e.g., Kp). This approach leads to a time varying Kp prediction and for those estimates to have quantifiable uncertainties.

In order to create this proof-of-concept, several assumptions and simplifications have been made. This is both a strength for the technique being computationally fast, as well as a weakness for the simplifications being unable to always capture the detailed nature of a complicated geomagnetic storm. A detailed statistical investigation is therefore required to further understand the probability distribution of accurate forecasts versus false positives. 

\begin{figure}
	\includegraphics[width=35pc]{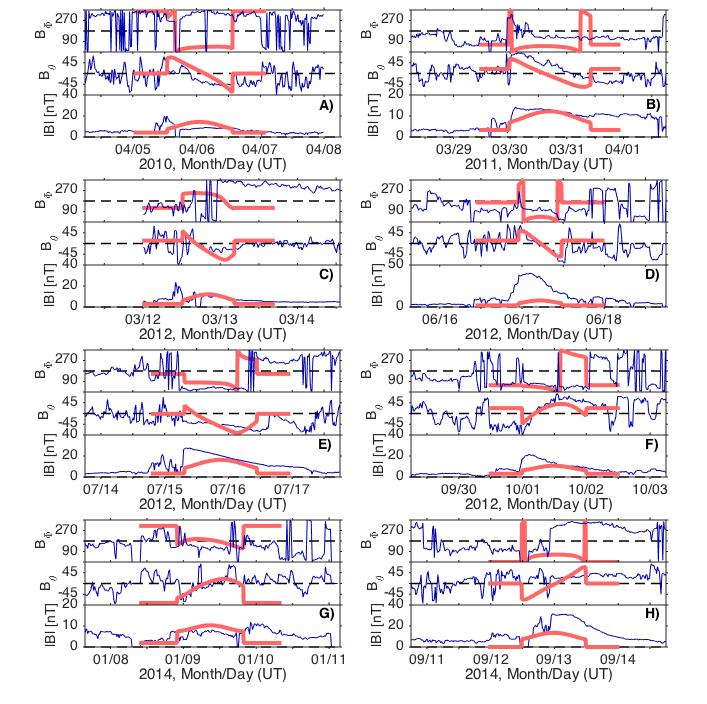} 
	\caption{Predicted (red) and observed (L1; blue) magnetic vectors for 8 CME events, where $\textbf{B}_{\theta}$ is the angular magnetic field direction out of the Sun-Earth plane.}
	\label{Bsurvey}
\end{figure}

The compressed solar wind plasma in between supersonic magnetic flux rope obstacles and their driven shock fronts has not been addressed in this article, even though they have been shown to be significant drivers of magnetospheric storms \citep{huttunen2004}. Panel C, D and E in Figure \ref{Bsurvey} displays the more pronounced examples of high amplitude fluctuations in the magnetic field just prior to the start of the flux rope CME. Therefore, future work may consider approaches that can better forecast these components of geo-effective CMEs.

In order to determine the usefulness of predicting the magnetic vectors for the purposes of estimating geomagnetic indices, a standardized procedure that all future techniques can be tested against will be beneficial. Such a forecast skill score (e.g., Heidke or Brier skill score) will perhaps be more useful than a traditional RMSE of individual data points between predicted and observed \citep[e.g.,][]{mays2015}, as this will potentially prevent uncertainty in arrival time values skewing the results.

\begin{acknowledgments}
This work was supported by NASA grant NNH14AX40I and NASA contract S-136361-Y to NRL. We thank Y-M Wang (NRL) for constructive comments about active region helicity, and M. Stockman (SWPC) and B. Murtagh (SWPC) for clarifying the forecasting policy and procedures at SWPC.
The OMNI data were obtained from the GSFC/SPDF OMNIWeb interface at http://omniweb.gsfc.nasa.gov
\end{acknowledgments}


\begin{thebibliography}{58}
	\providecommand{\natexlab}[1]{#1}
	\expandafter\ifx\csname urlstyle\endcsname\relax
	\providecommand{\doi}[1]{doi:\discretionary{}{}{}#1}\else
	\providecommand{\doi}{doi:\discretionary{}{}{}\begingroup
		\urlstyle{rm}\Url}\fi
	
	\bibitem[{\textit{{Al-Haddad} et~al.}(2012)\textit{{Al-Haddad},
			{Nieves-Chinchilla}, {Savani}, {M{\"o}stl}, {Marubashi}, {Hidalgo},
			{Roussev}, {Poedts}, and {Farrugia}}}]{alhaddad2012}
	{Al-Haddad}, N., T.~{Nieves-Chinchilla}, N.~P. {Savani}, C.~{M{\"o}stl},
	K.~{Marubashi}, M.~{Hidalgo}, I.~I. {Roussev}, S.~{Poedts}, and C.~J.
	{Farrugia} (2012), {Magnetic Field Configuration Models and Reconstruction
		Methods for Interplanetary Coronal Mass Ejections}, \textit{ArXiv e-prints}.
	
	\bibitem[{\textit{{Bothmer} and {Schwenn}}(1998)}]{bothmer1998}
	{Bothmer}, V., and R.~{Schwenn} (1998), {The structure and origin of magnetic
		clouds in the solar wind}, \textit{Annales Geophysicae}, \textit{16}, 1--24,
	\doi{10.1007/s00585-997-0001-x}.
	
	\bibitem[{\textit{{Brueckner} et~al.}(1995)\textit{{Brueckner}, {Howard},
			{Koomen}, {Korendyke}, {Michels}, {Moses}, {Socker}, {Dere}, {Lamy},
			{Llebaria}, {Bout}, {Schwenn}, {Simnett}, {Bedford}, and
			{Eyles}}}]{brueckner1995}
	{Brueckner}, G.~E., R.~A. {Howard}, M.~J. {Koomen}, C.~M. {Korendyke}, D.~J.
	{Michels}, J.~D. {Moses}, D.~G. {Socker}, K.~P. {Dere}, P.~L. {Lamy},
	A.~{Llebaria}, M.~V. {Bout}, R.~{Schwenn}, G.~M. {Simnett}, D.~K. {Bedford},
	and C.~J. {Eyles} (1995), {The Large Angle Spectroscopic Coronagraph
		(LASCO)}, \textit{Solar Physics}, \textit{162}, 357--402,
	\doi{10.1007/BF00733434}.
	
	\bibitem[{\textit{{Burlaga}}(1988)}]{burlaga1988}
	{Burlaga}, L.~F. (1988), {Magnetic clouds and force-free fields with constant
		alpha}, \textit{\jgr}, \textit{93}, 7217--7224,
	\doi{10.1029/JA093iA07p07217}.
	
	\bibitem[{\textit{{Colaninno} et~al.}(2013)\textit{{Colaninno}, {Vourlidas},
			and {Wu}}}]{colaninno2013}
	{Colaninno}, R.~C., A.~{Vourlidas}, and C.~C. {Wu} (2013), {Quantitative
		comparison of methods for predicting the arrival of coronal mass ejections at
		Earth based on multiview imaging}, \textit{Journal of Geophysical Research
		(Space Physics)}, \textit{118}, 6866--6879, \doi{10.1002/2013JA019205}.
	
	\bibitem[{\textit{{Cremades} and {Bothmer}}(2004)}]{cremades2004}
	{Cremades}, H., and V.~{Bothmer} (2004), {On the three-dimensional
		configuration of coronal mass ejections}, \textit{\aap}, \textit{422},
	307--322, \doi{10.1051/0004-6361:20035776}.
	
	\bibitem[{\textit{{D{\'e}moulin} and {Dasso}}(2009)}]{demoulin2009}
	{D{\'e}moulin}, P., and S.~{Dasso} (2009), {Magnetic cloud models with bent and
		oblate cross-section boundaries}, \textit{\aap}, \textit{507}, 969--980,
	\doi{10.1051/0004-6361/200912645}.
	
	\bibitem[{\textit{{D{\'e}moulin} et~al.}(2013)\textit{{D{\'e}moulin}, {Dasso},
			and {Janvier}}}]{demoulin2013}
	{D{\'e}moulin}, P., S.~{Dasso}, and M.~{Janvier} (2013), {Does spacecraft
		trajectory strongly affect detection of magnetic clouds?}, \textit{\aap},
	\textit{550}, A3, \doi{10.1051/0004-6361/201220535}.
	
	\bibitem[{\textit{{Dungey}}(1961)}]{dungey1961}
	{Dungey}, J.~W. (1961), {Interplanetary Magnetic Field and the Auroral Zones},
	\textit{Physical Review Letters}, \textit{6}, 47--48,
	\doi{10.1103/PhysRevLett.6.47}.
	
	\bibitem[{\textit{{Hale}}(1925)}]{hale1925}
	{Hale}, G.~E. (1925), {A Test of the Electromagnetic Theory of the Hydrogen
		Vortices Surrounding Sun-Spots}, \textit{Proceedings of the National Academy
		of Science}, \textit{11}, 691--696, \doi{10.1073/pnas.11.11.691}.
	
	\bibitem[{\textit{{Harra} et~al.}(2007)\textit{{Harra}, {Crooker}, {Mandrini},
			{van Driel-Gesztelyi}, {Dasso}, {Wang}, {Elliott}, {Attrill}, {Jackson}, and
			{Bisi}}}]{harra2007}
	{Harra}, L.~K., N.~U. {Crooker}, C.~H. {Mandrini}, L.~{van Driel-Gesztelyi},
	S.~{Dasso}, J.~{Wang}, H.~{Elliott}, G.~{Attrill}, B.~V. {Jackson}, and M.~M.
	{Bisi} (2007), {How Does Large Flaring Activity from the Same Active Region
		Produce Oppositely Directed Magnetic Clouds?}, \textit{Solar Physics},
	\textit{244}, 95--114, \doi{10.1007/s11207-007-9002-x}.
	
	\bibitem[{\textit{{Hidalgo} et~al.}(2002)\textit{{Hidalgo},
			{Nieves-Chinchilla}, and {Cid}}}]{hidalgo2002}
	{Hidalgo}, M.~A., T.~{Nieves-Chinchilla}, and C.~{Cid} (2002), {Elliptical
		cross-section model for the magnetic topology of magnetic clouds},
	\textit{\grl}, \textit{29}(13), 1637, \doi{10.1029/2001GL013875}.
	
	\bibitem[{\textit{{Howard} et~al.}(2008)\textit{{Howard}, {Moses}, {Vourlidas},
			{Newmark}, {Socker}, {Plunkett}, {Korendyke}, {Cook}, {Hurley}, {Davila},
			{Thompson}, {St Cyr}, {Mentzell}, {Mehalick}, {Lemen}, {Wuelser}, {Duncan},
			{Tarbell}, {Wolfson}, {Moore}, {Harrison}, {Waltham}, {Lang}, {Davis},
			{Eyles}, {Mapson-Menard}, {Simnett}, {Halain}, {Defise}, {Mazy}, {Rochus},
			{Mercier}, {Ravet}, {Delmotte}, {Auchere}, {Delaboudiniere}, {Bothmer},
			{Deutsch}, {Wang}, {Rich}, {Cooper}, {Stephens}, {Maahs}, {Baugh},
			{McMullin}, and {Carter}}}]{howard2008}
	{Howard}, R.~A., J.~D. {Moses}, A.~{Vourlidas}, J.~S. {Newmark}, D.~G.
	{Socker}, S.~P. {Plunkett}, C.~M. {Korendyke}, J.~W. {Cook}, A.~{Hurley},
	J.~M. {Davila}, W.~T. {Thompson}, O.~C. {St Cyr}, E.~{Mentzell},
	K.~{Mehalick}, J.~R. {Lemen}, J.~P. {Wuelser}, D.~W. {Duncan}, T.~D.
	{Tarbell}, C.~J. {Wolfson}, A.~{Moore}, R.~A. {Harrison}, N.~R. {Waltham},
	J.~{Lang}, C.~J. {Davis}, C.~J. {Eyles}, H.~{Mapson-Menard}, G.~M. {Simnett},
	J.~P. {Halain}, J.~M. {Defise}, E.~{Mazy}, P.~{Rochus}, R.~{Mercier}, M.~F.
	{Ravet}, F.~{Delmotte}, F.~{Auchere}, J.~P. {Delaboudiniere}, V.~{Bothmer},
	W.~{Deutsch}, D.~{Wang}, N.~{Rich}, S.~{Cooper}, V.~{Stephens}, G.~{Maahs},
	R.~{Baugh}, D.~{McMullin}, and T.~{Carter} (2008), {Sun Earth Connection
		Coronal and Heliospheric Investigation (SECCHI)}, \textit{Space Science
		Review}, \textit{136}, 67--115, \doi{10.1007/s11214-008-9341-4}.
	
	\bibitem[{\textit{{Howard} and {DeForest}}(2012)}]{howardt2012b}
	{Howard}, T.~A., and C.~E. {DeForest} (2012), {Inner Heliospheric Flux Rope
		Evolution via Imaging of Coronal Mass Ejections}, \textit{\apj},
	\textit{746}, 64, \doi{10.1088/0004-637X/746/1/64}.
	
	\bibitem[{\textit{{Huttunen} and {Koskinen}}(2004)}]{huttunen2004}
	{Huttunen}, K., and H.~{Koskinen} (2004), {Importance of post-shock streams and
		sheath region as drivers of intense magnetospheric storms and high-latitude
		activity}, \textit{Annales Geophysicae}, \textit{22}, 1729--1738,
	\doi{10.5194/angeo-22-1729-2004}.
	
	\bibitem[{\textit{{Janvier} et~al.}(2013)\textit{{Janvier}, {D{\'e}moulin}, and
			{Dasso}}}]{janvier2013}
	{Janvier}, M., P.~{D{\'e}moulin}, and S.~{Dasso} (2013), {Global axis shape of
		magnetic clouds deduced from the distribution of their local axis
		orientation}, \textit{\aap}, \textit{556}, A50,
	\doi{10.1051/0004-6361/201321442}.
	
	\bibitem[{\textit{{Kunkel} and {Chen}}(2010)}]{kunkel2010}
	{Kunkel}, V., and J.~{Chen} (2010), {Evolution of a Coronal Mass Ejection and
		its Magnetic Field in Interplanetary Space}, \textit{\apjl}, \textit{715},
	L80--L83, \doi{10.1088/2041-8205/715/2/L80}.
	
	\bibitem[{\textit{{Lemen} et~al.}(2012)\textit{{Lemen}, {Title}, {Akin},
			{Boerner}, {Chou}, {Drake}, {Duncan}, {Edwards}, {Friedlaender}, {Heyman},
			{Hurlburt}, {Katz}, {Kushner}, {Levay}, {Lindgren}, {Mathur}, {McFeaters},
			{Mitchell}, {Rehse}, {Schrijver}, {Springer}, {Stern}, {Tarbell}, {Wuelser},
			{Wolfson}, {Yanari}, {Bookbinder}, {Cheimets}, {Caldwell}, {Deluca}, {Gates},
			{Golub}, {Park}, {Podgorski}, {Bush}, {Scherrer}, {Gummin}, {Smith}, {Auker},
			{Jerram}, {Pool}, {Soufli}, {Windt}, {Beardsley}, {Clapp}, {Lang}, and
			{Waltham}}}]{Lemen2012}
	{Lemen}, J.~R., A.~M. {Title}, D.~J. {Akin}, P.~F. {Boerner}, C.~{Chou}, J.~F.
	{Drake}, D.~W. {Duncan}, C.~G. {Edwards}, F.~M. {Friedlaender}, G.~F.
	{Heyman}, N.~E. {Hurlburt}, N.~L. {Katz}, G.~D. {Kushner}, M.~{Levay}, R.~W.
	{Lindgren}, D.~P. {Mathur}, E.~L. {McFeaters}, S.~{Mitchell}, R.~A. {Rehse},
	C.~J. {Schrijver}, L.~A. {Springer}, R.~A. {Stern}, T.~D. {Tarbell}, J.-P.
	{Wuelser}, C.~J. {Wolfson}, C.~{Yanari}, J.~A. {Bookbinder}, P.~N.
	{Cheimets}, D.~{Caldwell}, E.~E. {Deluca}, R.~{Gates}, L.~{Golub}, S.~{Park},
	W.~A. {Podgorski}, R.~I. {Bush}, P.~H. {Scherrer}, M.~A. {Gummin},
	P.~{Smith}, G.~{Auker}, P.~{Jerram}, P.~{Pool}, R.~{Soufli}, D.~L. {Windt},
	S.~{Beardsley}, M.~{Clapp}, J.~{Lang}, and N.~{Waltham} (2012), {The
		Atmospheric Imaging Assembly (AIA) on the Solar Dynamics Observatory (SDO)},
	\textit{Solar Physics}, \textit{275}, 17--40,
	\doi{10.1007/s11207-011-9776-8}.
	
	\bibitem[{\textit{{Lepping} and {Wu}}(2010)}]{lepping2010}
	{Lepping}, R.~P., and C.-C. {Wu} (2010), {Selection effects in identifying
		magnetic clouds and the importance of the closest approach parameter},
	\textit{Annales Geophysicae}, \textit{28}, 1539--1552,
	\doi{10.5194/angeo-28-1539-2010}.
	
	\bibitem[{\textit{{Lepping} et~al.}(2006)\textit{{Lepping}, {Berdichevsky},
			{Wu}, {Szabo}, {Narock}, {Mariani}, {Lazarus}, and {Quivers}}}]{lepping2006}
	{Lepping}, R.~P., D.~B. {Berdichevsky}, C.-C. {Wu}, A.~{Szabo}, T.~{Narock},
	F.~{Mariani}, A.~J. {Lazarus}, and A.~J. {Quivers} (2006), {A summary of WIND
		magnetic clouds for years 1995-2003: model-fitted parameters, associated
		errors and classifications}, \textit{Annales Geophysicae}, \textit{24},
	215--245, \doi{10.5194/angeo-24-215-2006}.
	
	\bibitem[{\textit{{Liu} et~al.}(2008)\textit{{Liu}, {Luhmann}, {Huttunen},
			{Lin}, {Bale}, {Russell}, and {Galvin}}}]{liu2008}
	{Liu}, Y., J.~G. {Luhmann}, K.~E.~J. {Huttunen}, R.~P. {Lin}, S.~D. {Bale},
	C.~T. {Russell}, and A.~B. {Galvin} (2008), {Reconstruction of the 2007 May
		22 Magnetic Cloud: How Much Can We Trust the Flux-Rope Geometry of CMEs?},
	\textit{\apjl}, \textit{677}, L133--L136, \doi{10.1086/587839}.
	
	\bibitem[{\textit{{Liu} et~al.}(2010{\natexlab{a}})\textit{{Liu}, {Davies},
			{Luhmann}, {Vourlidas}, {Bale}, and {Lin}}}]{liu2010}
	{Liu}, Y., J.~A. {Davies}, J.~G. {Luhmann}, A.~{Vourlidas}, S.~D. {Bale}, and
	R.~P. {Lin} (2010{\natexlab{a}}), {Geometric Triangulation of Imaging
		Observations to Track Coronal Mass Ejections Continuously Out to 1 AU},
	\textit{\apjl}, \textit{710}, L82--L87, \doi{10.1088/2041-8205/710/1/L82}.
	
	\bibitem[{\textit{{Liu} et~al.}(2010{\natexlab{b}})\textit{{Liu}, {Thernisien},
			{Luhmann}, {Vourlidas}, {Davies}, {Lin}, and {Bale}}}]{liu2010b}
	{Liu}, Y., A.~{Thernisien}, J.~G. {Luhmann}, A.~{Vourlidas}, J.~A. {Davies},
	R.~P. {Lin}, and S.~D. {Bale} (2010{\natexlab{b}}), {Reconstructing Coronal
		Mass Ejections with Coordinated Imaging and in Situ Observations: Global
		Structure, Kinematics, and Implications for Space Weather Forecasting},
	\textit{\apj}, \textit{722}, 1762--1777, \doi{10.1088/0004-637X/722/2/1762}.
	
	\bibitem[{\textit{{Longcope} et~al.}(2007)\textit{{Longcope}, {Beveridge},
			{Qiu}, {Ravindra}, {Barnes}, and {Dasso}}}]{longcope2007}
	{Longcope}, D., C.~{Beveridge}, J.~{Qiu}, B.~{Ravindra}, G.~{Barnes}, and
	S.~{Dasso} (2007), {Modeling and Measuring the Flux Reconnected and Ejected
		by the Two-Ribbon Flare/CME Event on 7 November 2004}, \textit{Solar
		Physics}, \textit{244}, 45--73, \doi{10.1007/s11207-007-0330-7}.
	
	\bibitem[{\textit{{M{\"a}kel{\"a}} et~al.}(2013)\textit{{M{\"a}kel{\"a}},
			{Gopalswamy}, {Xie}, {Mohamed}, {Akiyama}, and {Yashiro}}}]{makela2013}
	{M{\"a}kel{\"a}}, P., N.~{Gopalswamy}, H.~{Xie}, A.~A. {Mohamed}, S.~{Akiyama},
	and S.~{Yashiro} (2013), {Coronal Hole Influence on the Observed Structure of
		Interplanetary CMEs}, \textit{Solar Physics}, \textit{284}, 59--75,
	\doi{10.1007/s11207-012-0211-6}.
	
	\bibitem[{\textit{{Manchester} et~al.}(2014)\textit{{Manchester}, {van der
				Holst}, and {Lavraud}}}]{manchester2014}
	{Manchester}, W.~B., IV, B.~{van der Holst}, and B.~{Lavraud} (2014), {Flux
		rope evolution in interplanetary coronal mass ejections: the 13 May 2005
		event}, \textit{Plasma Physics and Controlled Fusion}, \textit{56}(6),
	064006, \doi{10.1088/0741-3335/56/6/064006}.
	
	\bibitem[{\textit{Marubashi}(1997)}]{marubashi1997}
	Marubashi, K. (1997), Interplanetary magnetic flux ropes and solar filaments,
	in \textit{Coronal Mass Ejections}, \textit{Geophys. Monogr. Ser.}, vol.~99,
	edited by N.~{Crooker}, J.~A. {Joselyn}, and J.~{Feynman}, pp. 147--156, AGU,
	Washington, D. C., \doi{10.1029/GM099p0147}.
	
	\bibitem[{\textit{{Mays} et~al.}(2015)\textit{{Mays}, {Taktakishvili},
			{Pulkkinen}, {Odstrcil}, {MacNeice}, {Rastaetter}, {LaSota}, {Zheng}, and
			{Kuznetsova}}}]{mays2015}
	{Mays}, M.~L., A.~{Taktakishvili}, A.~A. {Pulkkinen}, D.~{Odstrcil}, P.~J.
	{MacNeice}, L.~{Rastaetter}, J.~A. {LaSota}, Y.~{Zheng}, and M.~M.
	{Kuznetsova} (2015), {Ensemble modeling of CMEs using the WSA-ENLIL+Cone
		model}, \textit{ArXiv e-prints}.
	
	\bibitem[{\textit{{McAllister} et~al.}(2001)\textit{{McAllister}, {Martin},
			{Crooker}, {Lepping}, and {Fitzenreiter}}}]{mcallister2001}
	{McAllister}, A.~H., S.~F. {Martin}, N.~U. {Crooker}, R.~P. {Lepping}, and
	R.~J. {Fitzenreiter} (2001), {A test of real-time prediction of magnetic
		cloud topology and geomagnetic storm occurrence from solar signatures},
	\textit{\jgr}, \textit{106}, 29,185--29,194, \doi{10.1029/2000JA000032}.
		
	\bibitem[{\textit{{M{\"o}stl} et~al.}(2014)\textit{{M{\"o}stl}, {Amla}, {Hall},
			{Liewer}, {De Jong}, {Colaninno}, {Veronig}, {Rollett}, {Temmer}, {Peinhart},
			{Davies}, {Lugaz}, {Liu}, {Farrugia}, {Luhmann}, {Vr{\v s}nak}, {Harrison},
			and {Galvin}}}]{mostl2014}
	{M{\"o}stl}, C., K.~{Amla}, J.~R. {Hall}, P.~C. {Liewer}, E.~M. {De Jong},
	R.~C. {Colaninno}, A.~M. {Veronig}, T.~{Rollett}, M.~{Temmer}, V.~{Peinhart},
	J.~A. {Davies}, N.~{Lugaz}, Y.~D. {Liu}, C.~J. {Farrugia}, J.~G. {Luhmann},
	B.~{Vr{\v s}nak}, R.~A. {Harrison}, and A.~B. {Galvin} (2014), {Connecting
		Speeds, Directions and Arrival Times of 22 Coronal Mass Ejections from the
		Sun to 1 AU}, \textit{\apj}, \textit{787}, 119,
	\doi{10.1088/0004-637X/787/2/119}.
	
	\bibitem[{\textit{{Nieves-Chinchilla} et~al.}(2012)\textit{{Nieves-Chinchilla},
			{Colaninno}, {Vourlidas}, {Szabo}, {Lepping}, {Boardsen}, {Anderson}, and
			{Korth}}}]{nieves2012}
	{Nieves-Chinchilla}, T., R.~{Colaninno}, A.~{Vourlidas}, A.~{Szabo}, R.~P.
	{Lepping}, S.~A. {Boardsen}, B.~J. {Anderson}, and H.~{Korth} (2012), {Remote
		and in situ observations of an unusual Earth-directed coronal mass ejection
		from multiple viewpoints}, \textit{Journal of Geophysical Research (Space
		Physics)}, \textit{117}, A06106, \doi{10.1029/2011JA017243}.
	
	\bibitem[{\textit{{Nieves-Chinchilla} et~al.}(2013)\textit{{Nieves-Chinchilla},
			{Vourlidas}, {Stenborg}, {Savani}, {Koval}, {Szabo}, and
			{Jian}}}]{nieves2013}
	{Nieves-Chinchilla}, T., A.~{Vourlidas}, G.~{Stenborg}, N.~P. {Savani},
	A.~{Koval}, A.~{Szabo}, and L.~K. {Jian} (2013), {Inner Heliospheric
		Evolution of a ''Stealth'' CME Derived from Multi-view Imaging and Multipoint
		in Situ observations. I. Propagation to 1 AU}, \textit{\apj}, \textit{779},
	55, \doi{10.1088/0004-637X/779/1/55}.
	
	\bibitem[{\textit{{Odstrcil} and {Pizzo}}(1999)}]{odstrcil1999}
	{Odstrcil}, D., and V.~J. {Pizzo} (1999), {Distortion of the interplanetary
		magnetic field by three-dimensional propagation of coronal mass ejections in
		a structured solar wind}, \textit{\jgr}, \textit{104}, 28,225--28,240,
	\doi{10.1029/1999JA900319}.
	
	\bibitem[{\textit{{Owens} and {Cargill}}(2004)}]{owens2004}
	{Owens}, M., and P.~{Cargill} (2004), {Predictions of the arrival time of
		Coronal Mass Ejections at 1AU: an analysis of the causes of errors},
	\textit{Annales Geophysicae}, \textit{22}, 661--671,
	\doi{10.5194/angeo-22-661-2004}.
	
	\bibitem[{\textit{{Owens} et~al.}(2006)\textit{{Owens}, {Merkin}, and
			{Riley}}}]{owens2006}
	{Owens}, M.~J., V.~G. {Merkin}, and P.~{Riley} (2006), {A kinematically
		distorted flux rope model for magnetic clouds}, \textit{Journal of
		Geophysical Research (Space Physics)}, \textit{111}, A03104,
	\doi{10.1029/2005JA011460}.
	
	\bibitem[{\textit{{Owens} et~al.}(2012)\textit{{Owens}, {D{\'e}moulin},
			{Savani}, {Lavraud}, and {Ruffenach}}}]{owens2012}
	{Owens}, M.~J., P.~{D{\'e}moulin}, N.~P. {Savani}, B.~{Lavraud}, and
	A.~{Ruffenach} (2012), {Implications of Non-cylindrical Flux Ropes for
		Magnetic Cloud Reconstruction Techniques and the Interpretation of Double
		Flux Rope Events}, \textit{Solar Physics}, \textit{278}, 435--446,
	\doi{10.1007/s11207-012-9939-2}.
	
	\bibitem[{\textit{{Poomvises} et~al.}(2012)\textit{{Poomvises}, {Gopalswamy},
			{Yashiro}, {Kwon}, and {Olmedo}}}]{poomvises2012}
	{Poomvises}, W., N.~{Gopalswamy}, S.~{Yashiro}, R.-Y. {Kwon}, and O.~{Olmedo}
	(2012), {Determination of the Heliospheric Radial Magnetic Field from the
		Standoff Distance of a CME-driven Shock Observed by the STEREO Spacecraft},
	\textit{\apj}, \textit{758}, 118, \doi{10.1088/0004-637X/758/2/118}.
	
	\bibitem[{\textit{{Richardson} and {Cane}}(2010)}]{richardson2010}
	{Richardson}, I.~G., and H.~V. {Cane} (2010), {Near-Earth Interplanetary
		Coronal Mass Ejections During Solar Cycle 23 (1996 - 2009): Catalog and
		Summary of Properties}, \textit{Solar Physics}, \textit{264}, 189--237,
	\doi{10.1007/s11207-010-9568-6}.
	
	\bibitem[{\textit{{Riley} and {Crooker}}(2004)}]{riley2004a}
	{Riley}, P., and N.~U. {Crooker} (2004), {Kinematic Treatment of Coronal Mass
		Ejection Evolution in the Solar Wind}, \textit{\apj}, \textit{600},
	1035--1042, \doi{10.1086/379974}.
	
	\bibitem[{\textit{{Rodriguez} et~al.}(2011)\textit{{Rodriguez}, {Mierla},
			{Zhukov}, {West}, and {Kilpua}}}]{rodriguez2011}
	{Rodriguez}, L., M.~{Mierla}, A.~N. {Zhukov}, M.~{West}, and E.~{Kilpua}
	(2011), {Linking Remote-Sensing and In Situ Observations of Coronal Mass
		Ejections Using STEREO}, \textit{Solar Physics}, \textit{270}, 561--573,
	\doi{10.1007/s11207-011-9784-8}.
	
	\bibitem[{\textit{{Ruffenach} et~al.}(2012)\textit{{Ruffenach}, {Lavraud},
			{Owens}, {Sauvaud}, {Savani}, {Rouillard}, {D{\'e}moulin}, {Foullon},
			{Opitz}, {Fedorov}, {Jacquey}, {G{\'e}not}, {Louarn}, {Luhmann}, {Russell},
			{Farrugia}, and {Galvin}}}]{ruffenach2012}
	{Ruffenach}, A., B.~{Lavraud}, M.~J. {Owens}, J.-A. {Sauvaud}, N.~P. {Savani},
	A.~P. {Rouillard}, P.~{D{\'e}moulin}, C.~{Foullon}, A.~{Opitz}, A.~{Fedorov},
	C.~J. {Jacquey}, V.~{G{\'e}not}, P.~{Louarn}, J.~G. {Luhmann}, C.~T.
	{Russell}, C.~J. {Farrugia}, and A.~B. {Galvin} (2012), {Multispacecraft
		observation of magnetic cloud erosion by magnetic reconnection during
		propagation}, \textit{Journal of Geophysical Research (Space Physics)},
	\textit{117}, A09101, \doi{10.1029/2012JA017624}.
	
	\bibitem[{\textit{{Savani} et~al.}(2010)\textit{{Savani}, {Owens}, {Rouillard},
			{Forsyth}, and {Davies}}}]{savani2010}
	{Savani}, N.~P., M.~J. {Owens}, A.~P. {Rouillard}, R.~J. {Forsyth}, and J.~A.
	{Davies} (2010), {Observational Evidence of a Coronal Mass Ejection
		Distortion Directly Attributable to a Structured Solar Wind}, \textit{\apjl},
	\textit{714}, L128--L132, \doi{10.1088/2041-8205/714/1/L128}.
	
	\bibitem[{\textit{{Savani} et~al.}(2011{\natexlab{a}})\textit{{Savani},
			{Owens}, {Rouillard}, {Forsyth}, {Kusano}, {Shiota}, and
			{Kataoka}}}]{savani2011a}
	{Savani}, N.~P., M.~J. {Owens}, A.~P. {Rouillard}, R.~J. {Forsyth},
	K.~{Kusano}, D.~{Shiota}, and R.~{Kataoka} (2011{\natexlab{a}}), {Evolution
		of Coronal Mass Ejection Morphology with Increasing Heliocentric Distance. I.
		Geometrical Analysis}, \textit{\apj}, \textit{731}, 109,
	\doi{10.1088/0004-637X/731/2/109}.
	
	\bibitem[{\textit{{Savani} et~al.}(2011{\natexlab{b}})\textit{{Savani},
			{Owens}, {Rouillard}, {Forsyth}, {Kusano}, {Shiota}, {Kataoka}, {Jian}, and
			{Bothmer}}}]{savani2011b}
	{Savani}, N.~P., M.~J. {Owens}, A.~P. {Rouillard}, R.~J. {Forsyth},
	K.~{Kusano}, D.~{Shiota}, R.~{Kataoka}, L.~{Jian}, and V.~{Bothmer}
	(2011{\natexlab{b}}), {Evolution of Coronal Mass Ejection Morphology with
		Increasing Heliocentric Distance. II. In Situ Observations}, \textit{\apj},
	\textit{732}, 117, \doi{10.1088/0004-637X/732/2/117}.
	
	\bibitem[{\textit{{Savani} et~al.}(2012)\textit{{Savani}, {Shiota}, {Kusano},
			{Vourlidas}, and {Lugaz}}}]{savani2012b}
	{Savani}, N.~P., D.~{Shiota}, K.~{Kusano}, A.~{Vourlidas}, and N.~{Lugaz}
	(2012), {A Study of the Heliocentric Dependence of Shock Standoff Distance
		and Geometry using 2.5D Magnetohydrodynamic Simulations of Coronal Mass
		Ejection Driven Shocks}, \textit{\apj}, \textit{759}, 103,
	\doi{10.1088/0004-637X/759/2/103}.
	
	\bibitem[{\textit{{Savani} et~al.}(2013{\natexlab{a}})\textit{{Savani},
			{Vourlidas}, {Shiota}, {Linton}, {Kusano}, {Lugaz}, and
			{Rouillard}}}]{savani2013b}
	{Savani}, N.~P., A.~{Vourlidas}, D.~{Shiota}, M.~G. {Linton}, K.~{Kusano},
	N.~{Lugaz}, and A.~P. {Rouillard} (2013{\natexlab{a}}), {A Plasma {$\beta$}
		Transition within a Propagating Flux Rope}, \textit{\apj}, \textit{779}, 142,
	\doi{10.1088/0004-637X/779/2/142}.
	
	\bibitem[{\textit{{Savani} et~al.}(2013{\natexlab{b}})\textit{{Savani},
			{Vourlidas}, {Pulkkinen}, {Nieves-Chinchilla}, {Lavraud}, and
			{Owens}}}]{savani2013a}
	{Savani}, N.~P., A.~{Vourlidas}, A.~{Pulkkinen}, T.~{Nieves-Chinchilla},
	B.~{Lavraud}, and M.~J. {Owens} (2013{\natexlab{b}}), {Tracking the momentum
		flux of a CME and quantifying its influence on geomagnetically induced
		currents at Earth}, \textit{Space Weather}, \textit{11}, 245--261,
	\doi{10.1002/swe.20038}.
	
	\bibitem[{\textit{{Shen} et~al.}(2012)\textit{{Shen}, {Wang}, {Wang}, {Liu},
			{Liu}, {Vourlidas}, {Miao}, {Ye}, {Liu}, and {Zhou}}}]{shen2012}
	{Shen}, C., Y.~{Wang}, S.~{Wang}, Y.~{Liu}, R.~{Liu}, A.~{Vourlidas},
	B.~{Miao}, P.~{Ye}, J.~{Liu}, and Z.~{Zhou} (2012), {Super-elastic collision
		of large-scale magnetized plasmoids in the heliosphere}, \textit{Nature
		Physics}, \textit{8}, 923--928, \doi{10.1038/nphys2440}.
	
	\bibitem[{\textit{{Shiota} et~al.}(2014)\textit{{Shiota}, {Kataoka}, {Miyoshi},
			{Hara}, {Tao}, {Masunaga}, {Futaana}, and {Terada}}}]{shiota2014}
	{Shiota}, D., R.~{Kataoka}, Y.~{Miyoshi}, T.~{Hara}, C.~{Tao}, K.~{Masunaga},
	Y.~{Futaana}, and N.~{Terada} (2014), {Inner heliosphere MHD modeling system
		applicable to space weather forecasting for the other planets}, \textit{Space
		Weather}, \textit{12}, 187--204, \doi{10.1002/2013SW000989}.
	
	\bibitem[{\textit{{Taktakishvili} et~al.}(2010)\textit{{Taktakishvili},
			{MacNeice}, and {Odstrcil}}}]{taktakishvili2010}
	{Taktakishvili}, A., P.~{MacNeice}, and D.~{Odstrcil} (2010), {Model
		uncertainties in predictions of arrival of coronal mass ejections at Earth
		orbit}, \textit{Space Weather}, \textit{8}, S06007,
	\doi{10.1029/2009SW000543}.
	
	\bibitem[{\textit{{Thernisien} et~al.}(2009)\textit{{Thernisien}, {Vourlidas},
			and {Howard}}}]{thernisien2009}
	{Thernisien}, A., A.~{Vourlidas}, and R.~A. {Howard} (2009), {Forward Modeling
		of Coronal Mass Ejections Using STEREO/SECCHI Data}, \textit{Solar Physics},
	\textit{256}, 111--130, \doi{10.1007/s11207-009-9346-5}.
	
	\bibitem[{\textit{{Tsurutani} and {Gonzalez}}(1997)}]{tsurutani1997}
	{Tsurutani}, B.~T., and W.~D. {Gonzalez} (1997), {The Interplanetary causes of
		magnetic storms: A review}, in \textit{Washington DC American Geophysical
		Union Geophysical Monograph Series}, \textit{Washington DC American
		Geophysical Union Geophysical Monograph Series}, vol.~98, edited by B.~T.
	{Tsurutani}, W.~D. {Gonzalez}, Y.~{Kamide}, and J.~K. {Arballo}, pp. 77--89,
	\doi{10.1029/GM098p0077}.
	
	\bibitem[{\textit{{Tucker-Hood} et~al.}(2015)\textit{{Tucker-Hood}, {Scott},
			{Owens}, {Jackson}, {Barnard}, {Davies}, {Crothers}, {Lintott}, {Simpson},
			{Savani}, {Wilkinson}, {Harder}, {Eriksson}, {L Baeten}, and {Wan
				Wah}}}]{tucker2015}
	{Tucker-Hood}, K., C.~{Scott}, M.~{Owens}, D.~{Jackson}, L.~{Barnard}, J.~A.
	{Davies}, S.~{Crothers}, C.~{Lintott}, R.~{Simpson}, N.~P. {Savani},
	J.~{Wilkinson}, B.~{Harder}, G.~M. {Eriksson}, E.~M. {L Baeten}, and L.~L.
	{Wan Wah} (2015), {Validation of a priori CME arrival predictions made using
		real-time heliospheric imager observations}, \textit{Space Weather},
	\textit{13}, 35--48, \doi{10.1002/2014SW001106}.
	
	\bibitem[{\textit{{Tun} and {Vourlidas}}(2013)}]{tun2013}
	{Tun}, S.~D., and A.~{Vourlidas} (2013), {Derivation of the Magnetic Field in a
		Coronal Mass Ejection Core via Multi-frequency Radio Imaging}, \textit{\apj},
	\textit{766}, 130, \doi{10.1088/0004-637X/766/2/130}.
	
	\bibitem[{\textit{{Vourlidas}}(2014)}]{vourlidas2014}
	{Vourlidas}, A. (2014), {The flux rope nature of coronal mass ejections},
	\textit{Plasma Physics and Controlled Fusion}, \textit{56}(6), 064001,
	\doi{10.1088/0741-3335/56/6/064001}.
	
	\bibitem[{\textit{{Vr{\v s}nak} et~al.}(2014)\textit{{Vr{\v s}nak}, {Temmer},
			{{\v Z}ic}, {Taktakishvili}, {Dumbovi{\'c}}, {M{\"o}stl}, {Veronig}, {Mays},
			and {Odstr{\v c}il}}}]{vrsnak2014}
	{Vr{\v s}nak}, B., M.~{Temmer}, T.~{{\v Z}ic}, A.~{Taktakishvili},
	M.~{Dumbovi{\'c}}, C.~{M{\"o}stl}, A.~M. {Veronig}, M.~L. {Mays}, and
	D.~{Odstr{\v c}il} (2014), {Heliospheric Propagation of Coronal Mass
		Ejections: Comparison of Numerical WSA-ENLIL+Cone Model and Analytical
		Drag-based Model}, \textit{\apjs}, \textit{213}, 21,
	\doi{10.1088/0067-0049/213/2/21}.
	
	\bibitem[{\textit{{Wang}}(2013)}]{wang2013}
	{Wang}, Y.-M. (2013), {On the Strength of the Hemispheric Rule and the Origin
		of Active-region Helicity}, \textit{\apjl}, \textit{775}, L46,
	\doi{10.1088/2041-8205/775/2/L46}.
	
	\bibitem[{\textit{{Yashiro} et~al.}(2004)\textit{{Yashiro}, {Gopalswamy},
			{Michalek}, {St.~Cyr}, {Plunkett}, {Rich}, and {Howard}}}]{yashiro2004}
	{Yashiro}, S., N.~{Gopalswamy}, G.~{Michalek}, O.~C. {St.~Cyr}, S.~P.
	{Plunkett}, N.~B. {Rich}, and R.~A. {Howard} (2004), {A catalog of white
		light coronal mass ejections observed by the SOHO spacecraft},
	\textit{Journal of Geophysical Research (Space Physics)}, \textit{109},
	A07105, \doi{10.1029/2003JA010282}.
	
	\bibitem[{\textit{{Zhang} and {Moldwin}}(2014)}]{zhang2014}
	{Zhang}, X.-Y., and M.~B. {Moldwin} (2014), {The source, statistical
		properties, and geoeffectiveness of long-duration southward interplanetary
		magnetic field intervals}, \textit{Journal of Geophysical Research (Space
		Physics)}, \textit{119}, 658--669, \doi{10.1002/2013JA018937}.
	
	\bibitem[{\textit{{Zheng} et~al.}(2013)\textit{{Zheng}, {Macneice}, {Odstrcil},
			{Mays}, {Rastaetter}, {Pulkkinen}, {Taktakishvili}, {Hesse}, {Masha
				Kuznetsova}, {Lee}, and {Chulaki}}}]{Zheng2013}
	{Zheng}, Y., P.~{Macneice}, D.~{Odstrcil}, M.~L. {Mays}, L.~{Rastaetter},
	A.~{Pulkkinen}, A.~{Taktakishvili}, M.~{Hesse}, M.~{Masha Kuznetsova},
	H.~{Lee}, and A.~{Chulaki} (2013), {Forecasting propagation and evolution of
		CMEs in an operational setting: What has been learned}, \textit{Space
		Weather}, \textit{11}, 557--574, \doi{10.1002/swe.20096}.
	
	\bibitem[{\textit{{Zurbuchen} and {Richardson}}(2006)}]{zurbuchen2006}
	{Zurbuchen}, T.~H., and I.~G. {Richardson} (2006), {In-Situ Solar Wind and
		Magnetic Field Signatures of Interplanetary Coronal Mass Ejections},
	\textit{Space Science Reviews}, \textit{123}, 31--43,
	\doi{10.1007/s11214-006-9010-4}.
	
\end{thebibliography}
%

\end{article}

%
\begin{figure}
\includegraphics[width=30pc]{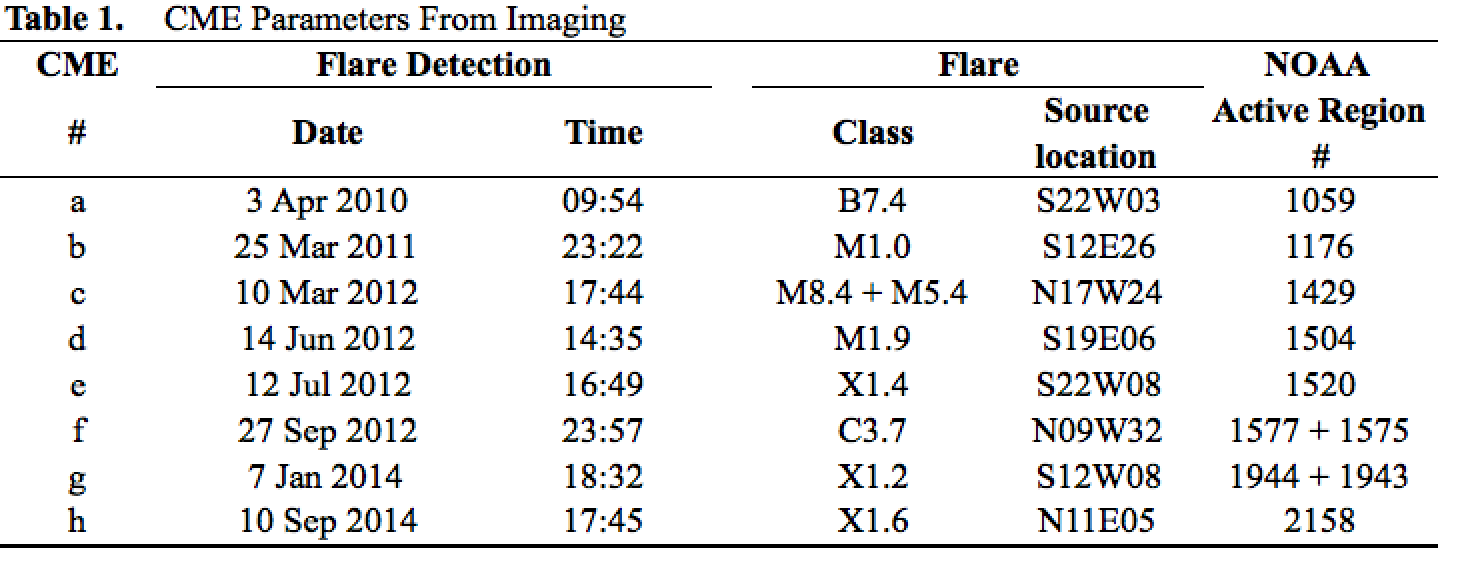}
	\label{table1}
\end{figure}

\end{document}